\documentclass[twocolumn,showpacs,pra,aps,superscriptaddress, amsmath, preprintnumbers,amssymb]{revtex4-2}
\usepackage[caption=false]{subfig}
\usepackage{graphicx} 
\usepackage{todonotes}
\usepackage{amsmath}
\usepackage{amssymb}
\usepackage{xspace}
\usepackage{orcidlink}
\usepackage{hyperref}
\usepackage{forest}
\usepackage[normalem]{ulem}

\usepackage{textcomp} 

\newcommand{\BNL}{Department of Physics, Brookhaven National Laboratory, Upton New York}

\newcommand{\VTQ}{Virginia Tech Center for Quantum Information Science and Engineering, Blacksburg, VA 24061}
\newcommand{\VTPhysics}{Department of Physics, Virginia Tech, Blacksburg, VA 24061}
\newcommand{\VTChemistry}{Department of Chemistry, Virginia Tech, Blacksburg, VA 24061}
\newcommand{\FSU}{Department of Physics, Florida State University, Tallahassee, FL 32306}
\newcommand{\Bielefeld}{Fakult\"at f\"ur Physik, Universit\"at Bielefeld, D-33615 Bielefeld, Germany}

\newcommand{\Adapt}{ADAPT-VQE\xspace}
\newcommand{\SCAdapt}{SC-ADAPT-VQE\xspace}

\newcommand{\vect}[1]{\boldsymbol{\mathbf{#1}}}

\newcommand{\Fig}[1]{Fig.\,\ref{fig:#1}} 
\newcommand{\Sec}[1]{Sec.\,\ref{sec:#1}} 

\newcommand{\easy}{$\xi_{A}$\xspace}
\newcommand{\medium}{$\xi_{B}$\xspace}
\newcommand{\hard}{$\xi_{C}$\xspace}

\newcommand{\REF}{{\rm ref}}

\newcommand{\Q}{Q\xspace}

\newcommand{\A}{$\Lambda$\xspace}
\newcommand{\T}{$\mathcal{T}$\xspace}
\newcommand{\Z}{Z\xspace}
\newcommand{\x}{$\ast$\xspace}

  \def\iu{\dot{\iota}} 

\newcommand{\xxx}{\x{}\x{}\x{}\xspace}
\newcommand{\xxZ}{\x{}\x{}\Z{}\xspace}

\newcommand{\xQx}{\x{}\Q{}\x{}\xspace}
\newcommand{\xQZ}{\x{}\Q{}\Z{}\xspace}
\newcommand{\Axx}{\A{}\x{}\x{}\xspace}
\newcommand{\AxZ}{\A{}\x{}\Z{}\xspace}

\newcommand{\AQx}{\A{}\Q{}\x{}\xspace}
\newcommand{\AQZ}{\A{}\Q{}\Z{}\xspace}

\newcommand{\Paulimm}{\boxplus}
\newcommand{\Qtiledmm}{{\boxplus_Q}} 
\newcommand{\Atiledmm}{{\boxplus_\Lambda}} 
\newcommand{\Pauli}{$\Paulimm$\xspace}
\newcommand{\Qtiled}{$\Qtiledmm$\xspace} 
\newcommand{\Atiled}{$\Atiledmm$\xspace} 

\begin{document}

\begin{abstract}
We investigate the role of symmetries in constructing resource-efficient operator pools for adaptive variational quantum eigensolvers. In particular, we focus on the lattice Schwinger model, a discretized model of $1+1$ dimensional electrodynamics, which we use as a proxy for spin chains with a continuum limit. We present an extensive set of simulations comprising a total of $11$ different operator pools, which all systematically and independently break or preserve a combination of discrete translations, the conservation of charge (``magnetization'') and the fermionic locality of the excitations. Circuit depths are the primary bottleneck in current quantum hardware, and we find that the most efficient ans\"atze in the near-term are obtained by pools that \textit{break} translation invariance, conserve charge, and lead to shallow circuits. On the other hand, we anticipate the shot counts to be the limiting factor in future, error-corrected quantum devices; our findings suggest that pools \textit{preserving} translation invariance could be preferable for such platforms.
\end{abstract}

\title{To break, or not to break: Symmetries in adaptive quantum simulations, a case study on the Schwinger model}

\author{Karunya Shirali${}^{\orcidlink{0000-0002-2006-2343}}$}
\email{karunyashirali@vt.edu}
\affiliation{\VTPhysics} \affiliation{\VTQ}

\author{Kyle Sherbert${}^{\orcidlink{0000-0002-5258-6539}}$}
\altaffiliation[Present Address: ]{Holy Cross College, Notre Dame, IN 46556}
\affiliation{\VTChemistry} \affiliation{\VTPhysics} \affiliation{\VTQ}

\author{Yanzhu Chen$^{\orcidlink{0000-0001-5589-9197}}$}
\affiliation{\FSU}
\affiliation{\VTPhysics} \affiliation{\VTQ}

\author{Adrien Florio${}^{\orcidlink{0000-0002-7276-4515}}$}
\affiliation{\Bielefeld}\affiliation{\BNL}

\author{Andreas Weichselbaum$^{\orcidlink{0000-0002-5832-3908}}\,$}
\affiliation{\BNL}

\author{Robert D. Pisarski${}^{\orcidlink{0000-0002-7862-4759}}$}
\affiliation{\BNL}

\author{Sophia E. Economou${}^{\orcidlink{0000-0002-1939-5589}}$}
\affiliation{\VTPhysics} \affiliation{\VTQ}

\date{\today}

\maketitle

\section{Introduction}

Quantum computing offers the prospect of tackling problems of particle physics such as lattice gauge theories that are intractable with classical computers, in particular, simulating the real time dynamics of the system as well as the properties of matter at finite density that are challenging due to the sign problem~\cite{Aarts_2016,Gattringer:2016kco,Alexandru:2020wrj,Troyer:2004ge}. 

Significant effort and discussion has been put forward for regularizing and digitizing the Hamiltonians for these gauge theories, mapping the infinite Hilbert space to finite resources. These methods include using finite groups \cite{Gustafson:2024kym,Bender:2018rdp,Hackett:2018cel,Alexandru:2019nsa,Yamamoto:2020eqi,Ji:2020kjk,Haase:2020kaj,Carena:2021ltu,Armon:2021uqr,Gonzalez-Cuadra:2022hxt,Charles:2023zbl,Irmejs:2022gwv,Gustafson:2020yfe, Bender:2018rdp,Hackett:2018cel,Alexandru:2019nsa,Yamamoto:2020eqi,Ji:2020kjk,Haase:2020kaj,Carena:2021ltu,Armon:2021uqr,Gonzalez-Cuadra:2022hxt,Charles:2023zbl,Irmejs:2022gwv, Hartung:2022hoz,Carena:2024dzu,Zohar:2014qma,Zohar:2016iic,Mueller:2024mmk}, q-deformed groups \cite{Zache:2023dko,Zache:2023cfj}, truncating representations \cite{Unmuth-Yockey:2018ugm,Unmuth-Yockey:2018xak, Klco:2019evd, Farrell:2023fgd, Farrell:2024fit, Ciavarella:2024lsp,Illa:2024kmf,Ciavarella:2021nmj, Bazavov:2015kka, Zhang:2018ufj,PhysRevD.99.114507,Bauer:2021gek,Grabowska:2022uos,Buser:2020uzs,Bhattacharya:2020gpm,Kavaki:2024ijd,Calajo:2024qrc,Murairi:2022zdg,Zohar:2015hwa,Zohar:2012xf,Zohar:2012ay,Zohar:2013zla}, quantum link models \cite{Brower:1997ha,Singh:2019jog,Singh:2019uwd,Wiese:2014rla,Brower:1997ha,Brower:2020huh,Mathis:2020fuo,Halimeh:2020xfd, budde2024quantum, osborne2024quantum, Osborne:2023rzx,Luo:2019vmi}, light front quantization \cite{Kreshchuk:2020dla,Kreshchuk:2020aiq,Kreshchuk:2020kcz}, loop-string-hadronization \cite{Davoudi:2024wyv,Raychowdhury:2018osk,Kadam:2023gli,Davoudi:2020yln,Mathew:2022nep}, and other methods \cite{Fromm:2023bit,Ciavarella:2024fzw,Alexandru:2023qzd}.
Less focus has been geared towards state preparation, which is an equally crucial problem. 

While Refs. \cite{Jordan:2011ci,Jordan:2011ne} pioneer using adiabatic evolution for state preparation, which has been a strong candidate \cite{DAnna:2024mmz,Chakraborty:2020uhf,Kaikov:2024acw}, this technique may not be amenable for theories such as quantum chromodynamics which exhibit multiple phase transitions. In this light other state preparation methods have been developed such as the density matrix method, E$\rho$ O q, \cite{Lamm:2018siq,Harmalkar:2020mpd,Gustafson:2020yfe} which uses Markov-Chain Monte-Carlo (MCMC) methods to identify states preparable with short depth circuits, cooling algorithms which iteratively drive heat out of the system using measurements \cite{Gustafson:2020vqg,Choi:2020tio,Lee:2019zze,Cohen:2023rhd,Qian:2021wya}, and more recently variational algorithms which utilize either a fixed ansatz or iteratively constructed ones to optimize a quantum target state \cite{Farrell:2024fit,Farrell:2023fgd,Gustafson2024:sc2,Nicoli:2025uzq,Zhang:2021bjq}.

The scalable circuit ADAPT-VQE method (SC-ADAPT-VQE) \cite{Farrell:2024fit,Farrell:2023fgd} and its generalization Surrogate Constructed Scalable Circuits ADAPT-VQE, (SC)$^2$-ADAPT-VQE \cite{Gustafson2024:sc2} are particularly novel developments among variational methods for lattice models. They utilize the framework of Adaptive Derivative-Assembled Problem Tailored (ADAPT)-VQE \cite{Grimsley:2018wnd} to identify a good ansatz and then use a multigrid approach of classical simulations to extrapolate parameters to a large volume system that can be simulated on a quantum computer.

The importance of symmetries and the effect of breaking them in VQE~\cite{Gard2020,Funcke2021dimensional,Ayeni2025,Gwangsu2025} and ADAPT-VQE simulations~\cite{Tsuchimochi2022,Bertels2022} has been addressed before in the context of spin and particle number conservation. The ans\"atze in \Adapt are constructed using operators from a user-defined \emph{pool} of operators, and early studies~\cite{Grimsley:2018wnd,Gyawali_2022_FermiHubbard}, including that in which the ADAPT-VQE algorithm was introduced, used operator pools based on fermionic single- and double-excitations. Fermionic single-excitations, for example, are operations that annihilate a particle in orbital/site $p$ and create a particle in orbital/site $q$, and double-excitations are their analog for pairs of particles. Subsequent work~\cite{Tang2021} demonstrated the qubit-ADAPT operator pool, that breaks down the Jordan-Wigner transformed fermionic excitations into individual Pauli terms, in addition to dropping the anti-commutation \Z strings, leading to operators that do not preserve the symmetries of the problem at hand. However, it is still able to accurately represent the ground-state, leading to drastically reduced circuit depths compared to the fermionic pool.
This suggests that it is not necessary to explicitly
conserve symmetries. However, other studies found that encoding the symmetries into the pool operators significantly improves the algorithm's convergence~\cite{Shkolnikov2021,Tsuchimochi2022}, and that breaking symmetries in the reference state slows down its convergence~\cite{Bertels2022}, suggesting that symmetries should be built into the simulations. Soon after, the qubit-excitation-based (QEB) pool~\cite{Yordanov2021} was shown to yield shallow circuit depths similar to those in qubit-ADAPT, thus retaining the hardware-efficiency as well as recovering particle-number conservation. More recently, the Coupled Exchange Operator operator pool~\cite{ramoa2024reducingresourcesrequiredadaptvqe} has been introduced and shown to have a highly compact circuit decomposition, while also preserving the particle-number and total Z spin projection ($S_Z$). 

Other symmetries such as translations, rotations, and reflections come into play in lattice models, which are ubiquitous in condensed matter and high energy physics. For example, in the Schwinger model, reflection, along with a simultaneous fermion-antifermion transformation, constitutes the parity symmetry. In Ref.~\cite{Farrell:2023fgd}, \SCAdapt was introduced and used to prepare the ground-state of the lattice Schwinger model with high fidelity. The authors also made use of translation-symmetry, constructing explicitly translation-invariant operators, and showed efficient circuit decompositions of their pool operators (albeit incurring some Trotterization error). Their choice of pool results in an (approximately) translation-invariant ansatz, that has well-defined extrapolations to large system volumes. 
This leads us to the question: what is the effect of breaking translation-symmetry in lattice models with additional symmetries using ADAPT-VQE? More specifically, 
\begin{enumerate}
    \item Is it more or less quantum-resource efficient?
    \item Does relaxing translation symmetry affect the accuracy of the optimized state?
\end{enumerate}
In addition, is there a distinction between discrete and continuous symmetries for \Adapt, or is the algorithm equally sensitive to the breaking of any symmetry? In the continuum limit, the conservation of charge (a discrete quantity) arises from the gauge field symmetry, which is continuous. Translation invariance is also a continuous symmetry in the continuum. Time-reversal symmetry, on the other hand, is a discrete symmetry. In lattice models, the gauge field symmetry is still continuous, leading to a (discrete) conserved charge, but translation invariance is discrete. 

\begin{figure}[t]
\label{fig:symmetries}
\includegraphics[width=\linewidth]{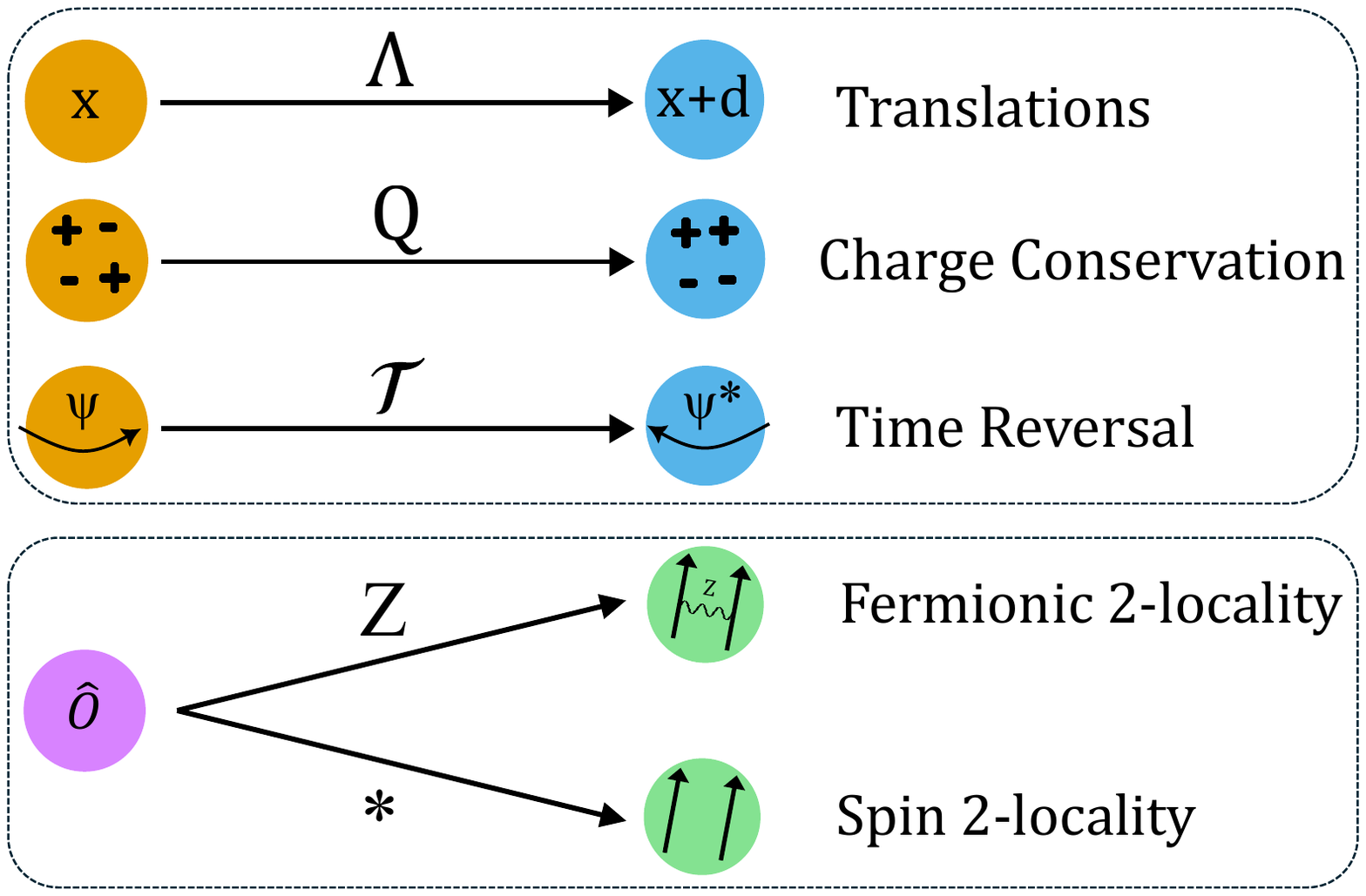}
\caption{Examples of various types of symmetries in particle physics and their effects on a given particle. Preservation of any of these symmetries means that they correspond to a conserved quantity, i.e. if the Hamiltonian is invariant under charge conjugation, it means that charges are conserved.
}
\end{figure}

In this work, we address this question by systematically and independently relaxing translation symmetry and charge conservation in the ansatz in adaptive variational simulations of the lattice Schwinger model, via appropriate choices of operator pools.
At the same time, we also consider the impact of retaining the strings of $Z$-Paulis arising from the Jordan-Wigner transformation.
Finally, we also perform experiments to relax the time-reversal symmetry. We assess the importance of each property in the operator pool by contrasting convergence rate, scalability, and circuit depth.

The manuscript is organized as follows: in Section \ref{sec:theory}, we describe the \Adapt algorithm and give an introduction to the lattice Schwinger model. In Section \ref{sec:pools}, we introduce the symmetry-relaxing operator pools used in this work. Section \ref{sec:results} contains our results and Section \ref{sec:discussion} includes a discussion of their implications.

\section{Theory}
\label{sec:theory}
\subsection{\Adapt}
The Adaptive Derivative-Assembled Problem Tailored Variational Quantum Eigensolver (ADAPT)-VQE is an algorithm that constructs ans\"{a}tze for ground-state preparation on the fly by applying layers of unitaries generated by problem-informed operators iteratively to a reference state. ADAPT-VQE builds the ansatz one step at time by choosing and adding to it an operator from a user-defined operator pool. The selection criterion used to choose the best operator at each step is a `greedy' strategy, where the algorithm chooses the operator that has the largest magnitude of the energy gradient. 

We define the pool of operators to be $\mathcal{O} = \{ \hat O^{(1)}, \hat O^{(2)}, \dots \hat O^{(N)}\}$. Then, the state at step $k$ of the algorithm can be expressed as
\begin{equation}
    |\psi_k (\vect{\theta}^{(k)})\rangle = \prod_{j=1}^k e^{-\iu \theta_j^{(k)} \hat{O}_{j}} |\psi_{\REF}\rangle,
\end{equation}
where $\vect{\theta}^{(k)}=(\theta_1^{(k)},\dots,\theta_k^{(k)})$, $\theta_j^{(k)}$ is the coefficient of operator $\hat{O}_{j}$ at step $k$, and $|\psi_{\REF}\rangle$ is the initial state.

The energy gradient of each pool operator $\hat O^{(i)}$ is then measured, which can be written as
\begin{equation}\label{adapt-gradient}
    G_i \equiv  \langle \psi_{k}(\vect{\theta}^{(k)}) | \left[ \hat O^{(i)}, \hat H \right] | \psi_{k}(\vect{\theta}^{(k)}) \rangle .
\end{equation}
Defining $\hat O_{k+1}$ to be the operator $\hat O^{(i)}$ with the largest-magnitude gradient $|G_i|$ in Eq.~\ref{adapt-gradient}, the ansatz is then grown using $\hat O_{k+1}$ to generate the new ansatz at step $k+1$ as 
\begin{equation}\label{adapt-vqe-eqn}
    |\psi_{k+1}(\vect{\theta}^{(k+1)})\rangle = e^{-\iu\theta_{k+1 }\hat O_{k+1}}|\psi_{k}(\vect{\theta}^{(k)})\rangle.
\end{equation}
After each step in which the ansatz is grown, the coefficients $\vect{\theta}$ are updated to minimize the energy. The process of growing the ansatz and optimizing coefficients is terminated when the largest $G_i$ falls below a threshold, $\varepsilon$. 

We note that the initial state $|\psi_{\REF}\rangle$ is a user-defined input to the algorithm, and is typically chosen so as to preserve the problem symmetries. Starting from an initial state with well-defined symmetries as in the ground state sector, and an operator pool that can break symmetries (or vice versa), leads to an ansatz in which the symmetry is broken in principle (but can be regained at a later iteration by applying an appropriate combination of operators). It is pertinent then to keep in mind that the symmetry of an ansatz is determined by both the initial state and the operator pool. For certain problems, if one wants to focus on a particular symmetry sector (say, a fixed baryon number), then ensuring that one starts and remains in the correct charge symmetry sector is important. 

\subsection{Lattice Schwinger model}
The Schwinger Hamiltonian is a model of quantum electrodynamics in $1+1D$ describing fermions coupled to a $U(1)$ gauge field.
We will study the model discretized using Kogut-Susskind or staggered lattice fermions \cite{PhysRevD.11.395}, using the Jordan-Wigner transformation to map fermionic sites onto qubits and Gauss's law to integrate out the gauge field links which has been widely used in quantum computing applications \cite{Gustafson:2023aai, Chakraborty:2020uhf,Honda:2021aum,Farrell:2023fgd,Farrell:2024fit,Farrell:2024fit,Osborne:2023rzx}.

The resulting Hamiltonian is
\begin{align}
\label{eq:schwingerham}
    \hat{H} = &\frac{1}{2a} \sum_{j=0}^{2L - 2} (\hat{\sigma}^+_j \hat{\sigma}^-_{j+1} + \hat{\sigma}^+_{j+1}\hat{\sigma}^-_{j})+ \frac{m_0}{2} \sum_{j=0}^{2L-1}(-1)^j\hat{\sigma}^z_j\notag\\
    & + \frac{a g^2}{8} \sum_{j=0}^{2L - 2} (\sum_{k=0}^{j}(\hat{\sigma}^z_k + (-1)^k))^2,
\end{align}
where $a$ is the lattice spacing,  $m_0$ is the bare mass of the fermion, $g$ is the coupling constant for the gauge field, and $L$ is the number of physical sites. The Kogut-Susskind fermions utilize a staggered lattice where the spin degrees of freedom are diagonalized so fermions live on even numbered sites and antifermions live on odd numbered sites. The ground state of the Schwinger model has charge $Q=0$.

\begin{figure}
    \centering
    \subfloat[\label{fig:schematic_relaxing_symmetries}]{
        \includegraphics[width=0.9\linewidth,trim={{0.9\linewidth} {.9\linewidth} {.9\linewidth} {.8\linewidth}},clip]{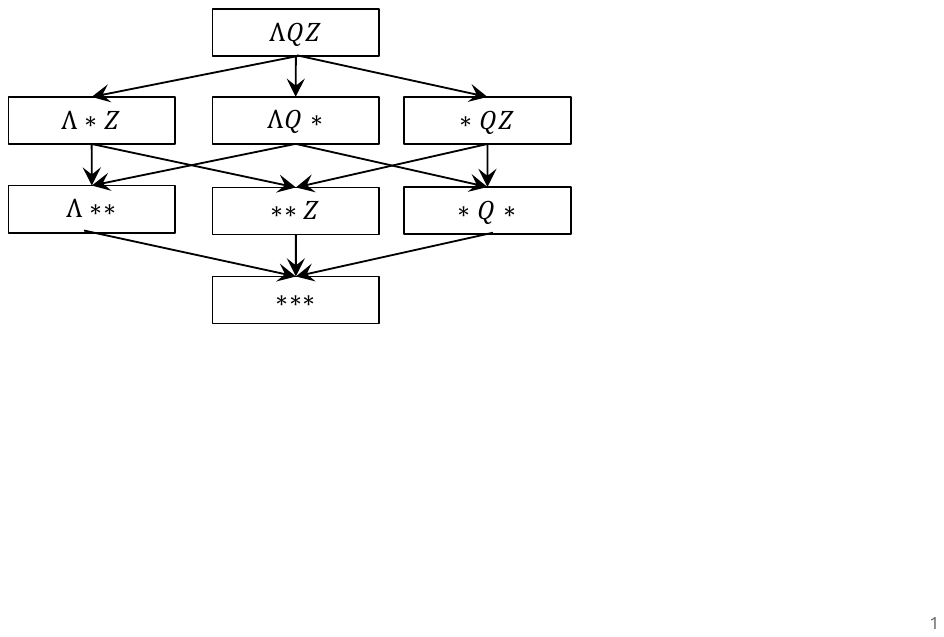}
    }\vspace{-0.4cm}
    \subfloat[\label{fig:bottom_up}]{
        \includegraphics[width=0.9\linewidth,trim={{\linewidth} {1.05\linewidth} {\linewidth} {1.1\linewidth}},clip]{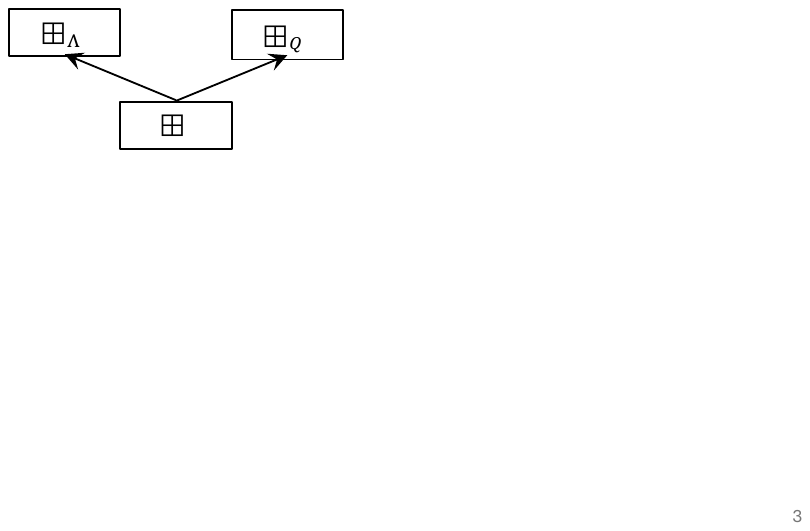}
    }
    \caption{Schematic depicting the hierarchy of relaxing symmetries in the operator pools in this work for \Adapt. The pools shown here all preserve time-reversal \T.
        (a) The pools at the top level preserve every symmetry in the problem Hamiltonian, and we relax these symmetries to form new pools, using a top-down approach. 
        (b) The \Atiled and \Qtiled pools are formed using a different, bottom-up approach, in which elementary tiled Pauli operators are used to construct more complex operators conserving translation invariance and charge.
    }
    \label{pool_hierarchies}
\end{figure}

\section{Operator Pools}
\label{sec:pools}
The goal of this paper is to study the accuracy and convergence of ADAPT-VQE when using operator pools that incorporate different symmetries.
As we indicate in Fig.~\ref{fig:schematic_relaxing_symmetries}, we can take a top-down approach, where we start with a pool that preserves every symmetry in the problem Hamiltonian, and we form new pools by relaxing each symmetry one by one.
Alternatively, we can take a bottom-up approach as in Fig.~\ref{fig:bottom_up}, using the operator tiling framework to build up increasingly complex pools incorporating new symmetries.

\subsection{Top-down Pool Selection}
\label{sec:topdown}
For the top of our hierarchy, we take the pool presented in \cite{Farrell:2023fgd}, which has several properties we label with the letters \A, \Q, \Z, and \T (cf. \Fig{symmetries}),
\begin{enumerate}
    \item Coordinate invariance (\A)
    \item Charge conservation (\Q)
    \item Fermionic 2-locality (\Z)
    \item Time reversal symmetry (\T)
\end{enumerate}
Properties \A, \Q, and \T correspond to physical global symmetries preserved by the operator, while property \Z is a representation-specific property of the operator informing the physical interpretation of the resulting ansatz.

We relax each of these properties independently.
We  refer to the pool with all these properties as the \AQZ pool, and to pools relaxing each property by replacing the letter with an asterisk (\x).
For example, we refer to the pool of {\em site-specific} charge-conserving fermionic 2-local operators (equivalent to single-excitation Unitary Coupled Cluster~\cite{BARTLETT1989133,Anand2022} operators) as the \xQZ pool.
In the following sections, we describe each property and describe how it is relaxed; see Fig.~\ref{fig:schematic_relaxing_symmetries} for a schematic summary.
In Section~\ref{sec:topdown:T}, we discuss time reversal symmetry, which is treated separately from \A, \Q and \Z.

\subsubsection{Coordinate Invariance (\A)}
\label{sec:topdown:A}
Coordinate invariant operators are those which are not explicitly labeled by a particular site index.
Reference \cite{Farrell:2023fgd} divides such operators into two categories: {\em volume} (or bulk) operators and {\em surface} (or boundary) operators (see \Sec{TIPP} for examples).
Volume operators are constructed to be a translationally invariant sum of terms, such that every cyclic permutation of site indices within each term results in the same operator.
Surface operators are constructed to act only on or near the boundaries of the simulation space.
To simulate a system with open boundary conditions, surface operators are necessary to break translational invariance.
In addition to guaranteeing ans\"atze with (approximate) translational symmetry, coordinate invariance is an especially desirable property for pool operators in \Adapt, because it ensures the same sequence of operators is well-defined for any volume.
This enables efficient extrapolation procedures for both variational parameters and quantum observables \cite{Farrell:2023fgd,Gustafson2024:sc2} and lends itself to many other problems.

Even with open boundary conditions, surface operators become less and less relevant as the system increases in size.
Therefore, we consider this property to be equivalent to preserving translational symmetry for large volumes.
To relax coordinate invariance, we partition the volume operator into its action on individual sites, and we take each partition as its own independent pool operator.
For instance, a volume operator in \AQZ and its set of translation-broken offspring operators in \xQZ might look like
\small
\begin{align*}
    (\Lambda Q Z) \leftarrow &\frac{1}{2} (XZZYII - YZZXII + IIXZZY - IIYZZX \\
    &- IXZZYI + IYZZXI). \\
    (* Q Z) \leftarrow &\frac{1}{2} (XZZYII - YZZXII),\\ 
    &\frac{1}{2} (IIXZZY - IIYZZX), \\
    &\frac{1}{2} (IXZZYI - IYZZXI).
\end{align*}
\normalsize

We note that the \AQZ and \AQx pools are symmetric under charge-conjugation (C) with simultaneous parity under inversion (P), i.e., reflection around the mid-point of the lattice, as are a few operators in the \xQZ and \xQx pools. 
On the other hand, all operators in the \AxZ, \Axx, \xxZ and \xxx pools break CP. 

\subsubsection{Charge conservation (\Q)}
\label{sec:topdown:C}

A charge conserving operator is one which leaves the charge invariant when it acts on any state.
When using the Jordan-Wigner transformation, charge conserving operators conserve the Hamming weight of any configuration, i.e. the number of ones in the computational basis state, so that in each term, every raising operator $\sigma^+$ is accompanied by a lowering operator $\sigma_-$.
Since the operator must also be Hermitian, charge conserving operators consist of pairs of terms like $(\sigma_i^-\sigma_j^+ + \sigma_i^+\sigma_j^-)$ or $i(\sigma_i^-\sigma_j^+ - \sigma_i^+\sigma_j^-)$.
In the Pauli representation, raising and lowering operators have the following form:
\begin{align}
    \sigma^{\pm} &= \tfrac{1}{2}(X \mp iY).
\end{align}
Therefore, charge conserving operators have one of the following forms:
\begin{align}
    \label{eq:C_op}  (\sigma_i^-\sigma_j^+ + \sigma_i^+ \sigma_j^-) &= \tfrac{1}{2}(X_i X_j + Y_i Y_j) \\
    \label{eq:C_op:T} \iu(\sigma_i^-\sigma_j^+ - \sigma_i^+ \sigma_j^-) &= \tfrac{\iu}{2}(X_i Y_j - Y_i X_j).
\end{align}
Most operators used in this work 
are of type \eqref{eq:C_op:T},
as these respect time-reversal symmetry.

To relax charge conservation, we partition the linear combination of Paulis (e.g. $X_i Y_j$ and $Y_i X_j$ in Eq.~\ref{eq:C_op:T}) into two distinct pool operators.
The non-conservation of charge can be seen directly, by writing the individual Pauli terms from Eqs.~\ref{eq:C_op} and~\ref{eq:C_op:T} in terms of raising and lowering operators $\sigma^\pm$.
\begin{align}
    X_i X_j &= \hspace{1em} (\sigma_i^-\sigma_j^+ + \sigma_i^+\sigma_j^-) + \hspace{0.33em} (\sigma_i^- \sigma_j^- + \sigma_i^+ \sigma_j^+) \\
    Y_i\, Y_j\, &= \hspace{1em} (\sigma_i^-\sigma_j^+ + \sigma_i^+\sigma_j^-) - \hspace{0.33em} (\sigma_i^- \sigma_j^- + \sigma_i^+ \sigma_j^+) \\
    X_i Y_j\, &= \hspace{0.67em} \iu(\sigma_i^-\sigma_j^+ - \sigma_i^+\sigma_j^-) - \iu(\sigma_i^- \sigma_j^- - \sigma_i^+ \sigma_j^+) \\
    Y_i X_j &= -\iu(\sigma_i^-\sigma_j^+ - \sigma_i^+\sigma_j^-) - \iu(\sigma_i^- \sigma_j^- - \sigma_i^+ \sigma_j^+).
\end{align}
These include contributions from $\sigma_i^+ \sigma_j^+$ and $\sigma_i^- \sigma_j^-$, which explicitly add or remove Hamming weight, violating charge conservation.

\subsubsection{Fermionic 2-locality (\Z)}
\label{sec:topdown:Z}
Fermionic 2-local operators are those in which each term consists of just two fermionic operators, typically a single creation operator and a single annihilation operator.
The two operators can act on different fermionic modes, labeled $i$ and $j$.
These are also referred to as single-body operators, since they represent a hopping of a single fermion from one mode to another.
Ans\"atze constructed from fermionic 2-local operators can thus be interpreted as a simple series of excitations, moving fermions between modes.
We briefly note that ans\"{a}tze with exclusively single-body operators are classically simulable; nevertheless we adopt 2-local fermionic operators as our starting point for top-down pool selection for the sake of a simplified experimental framework.

For our operators to be implementable in a quantum computer, we must move from fermionic representation to Pauli representation.
We use the Jordan-Wigner transformation, which maps any single fermionic operator to the corresponding spin operator, together with the necessary Z strings to enforce fermionic antisymmetry under permutation.
Importantly, the presence of the Z string causes each 2-local fermionic term to be mapped onto a linear combination of $O(n)$-local Pauli operators.
For example,
\begin{equation}
    a_1^\dagger a_4 \xrightarrow{JW} \sigma_1^+ \otimes Z_2 Z_3 \otimes \sigma_4^-.
\end{equation}
Alternatively, we could design a new Pauli operator by omitting the Z string, recovering 2-locality in the Pauli representation.
Note that this does not violate fermionic antisymmetry; 
rather, our new Pauli operator corresponds to a different fermionic many-body operator
under the Jordan-Wigner transformation which is $O(n)$-local.
For example,
\begin{multline}
    \sigma_1^+ \sigma_4^- \xleftarrow{JW} a_1^\dagger a_4 - 2 (a_2^\dagger a_1^\dagger a_4 a_2 + a_3^\dagger a_1^\dagger a_4 a_3) \\
    - 4\, 
       a_3^\dagger a_2^\dagger a_1^\dagger a_4 a_3 a_2.
\end{multline}
Modifying pool operators in this way has been shown to maintain the accuracy of \Adapt, trading a small penalty to convergence for a significantly more compact quantum circuit decomposition \cite{Tang2021,Yordanov2021}.
Furthermore, using qubit-local pool operators enables several sophisticated modifications to the original \Adapt protocol which significantly reduce its measurement overhead \cite{Anastasiou2024:TETRIS,Anastasiou2023:measurement}.
Motivated by these results, we relax fermionic 2-locality in this work by omitting the Z strings resulting from the Jordan Wigner transformation.

\subsubsection{Time-reversal symmetry (\T)}\label{TRS}
\label{sec:topdown:T}

For a time-independent system, in the presence of time-reversal
symmetry, the Hamiltonian $H$ can be chosen purely real. Hence any eigenstate can be represented by  a real wave function
up to a global phase.  For a generic state, $\psi(t)$ and
$\psi^{*}(-t)$ obey identical time-dependent Schr\"{o}dinger
equations with the same preserved energy expectation value.
Yet, in general, they are intrinsically complex.  Nevertheless
with (approximate) eigenstates of a time-reversal symmetric
Hamiltonian in mind, we take the insistence on a real wave
function as synonymous for respecting time-reversal symmetry.
This is achieved by evolution with purely imaginary operators.

In \Adapt, for a time-reversal symmetric target state we thus
start from a real reference state $|\psi_{\REF}\rangle$ and only
include purely imaginary operators $\hat O$ in the pool.
This ensures that the unitaries $e^{\iu\theta \hat O}$ evolving the ansatz forward are real-valued and preserve wave-function reality. This can be fulfilled by using operators in which each Pauli string contains an odd number of $Y$'s. We note that for real reference states, operators that break \T (those that have an even number of $Y$'s), such as $XXYY$, are never chosen by \Adapt because the gradients ($\langle [H,O] \rangle$) of these operators are always zero. In particular, the expectation value (see Eq.~\ref{adapt-gradient}) of the commutator over real reference states is guaranteed to be zero.
This can be understood as follows: for \T-breaking operators $O$ and a Hamiltonian $H$ preserving \T, $[ H, O]$ always contains an odd number of $Y$'s, i.e., is entirely imaginary (but also, Hermitian). Now, the expectation value of a Hermitian operator must always be real; thus, for real reference states, enforcing that $\langle [ H, O ] \rangle $ must be real implies that it must be exactly zero. On the other hand, for complex reference states, $\langle [ H, O ] \rangle $ must still remain real, but can be non-zero.

Relaxing time-reversal symmetry for the simulations shown in this paper, therefore, requires introducing complex amplitudes into the reference state such that it cannot be written as a real wavefunction with an overall complex phase. In Sec.~\ref{TRS-breaking}, we show the results of simulations in which time-reversal symmetry is broken.

\subsection{Bottom-up Pool Selection}\label{sec:tiling}

\subsubsection{Operator tiling}\label{operator_tiling}
Recently, Ref.~\cite{OperatorTiling2024} introduced the operator pool tiling technique to construct linearly scaling operator pools for translation-invariant lattice models in which solutions of \Adapt for a small problem instance are used to build an operator pool for larger problem instances. The resultant ans\"{a}tze produced by the method are qubit-local, as well as restricted to a symmetry subsector of the Hilbert space that obey a subset of symmetries of the Hamiltonian (in this case, the $\mathbb{Z}_{2}$ symmetry corresponding to the fermion/anti-fermion number parities, and time-reversal symmetry). 

The general procedure to generate the tiled operators is as follows:
\begin{enumerate}
    \item\label{degenerateADAPT} Collect operators $\{ \Paulimm_j \}$ acting on $L_{\rm tile}$ qubits
    chosen from several \Adapt runs on a small problem instance using a highly expressive operator pool.
    \item Tile the operators chosen by \Adapt for the small problem instance, $\{ \Paulimm_j \}$, to larger problem instances to form a tiled operator pool, $\{ \bigcup_{i,j} \ I^{\otimes i}\otimes \Paulimm_j \otimes I^{\otimes (n-i-L_{\rm tile})}\}$.
\end{enumerate}
This standardized procedure lets \Adapt pick out problem-relevant local interactions, even if they are not obviously contained within the Hamiltonian.

There are various choices of operator pools that could be used for Step 1 above. Here, we use an operator pool consisting of all possible Pauli strings when running \Adapt on the small problem instance and tile the resulting set of operators to larger problem instances. Individual Paulis cannot take on the form of Eq.~\ref{eq:C_op:T} and hence break the charge conservation. At the same time, these operators are not coordinate invariant either, since they are not invariant under a cyclic permutation of site indices. Thus, this pool is at the bottom of the symmetry hierarchy in Fig.~\ref{fig:bottom_up}. 

In principle, a pool of all possible Pauli strings, $\{ \Paulimm^{\otimes n} \}$ constitutes an entirely problem-agnostic approach in which no a priori knowledge of symmetries is used in constructing the operator pool, and could be used as a pool. However, this is impractical for two reason: first, the full Pauli pool contains an exponential number of operators, rendering it unfavorable for current hardware; second, it contains many operators that violate the parity and time-reversal symmetries. As discussed in Sec.~\ref{TRS}, ADAPT-VQE never selects such operators that break the symmetry of the reference state. There are thus a large number of operators that are unnecessary to include in the operator pool. Therefore, we treat the tiled Pauli pool as the elementary pool from which more complex pools can be constructed.

\subsubsection{The charge-conserving Pauli Pool}
The tiled Pauli pool operators do not conserve charge. To further explore the importance of charge conservation, we design a set of operators that conserve charge, by constructing linear combinations of commuting Pauli operators chosen by \Adapt for the small problem instance. This pool, termed the \Qtiled (read as Q-tiled) pool, is arrived at using a bottom-up approach, by designing operators of greater complexity from simple Paulis. 

For example, consider the operators $Z_1 I_2 X_3 Y_4, \: I_1 Z_2 X_3 Y_4, \: Z_1 I_2 Y_3 X_4,\: I_1 Z_2 Y_3 X_4$. While they do not individually conserve charge, the operator
\begin{multline}\label{charge-conserving-1}
    \Qtiledmm =\frac{1}{4} (Z_1 I_2 X_3 Y_4 - I_1 Z_2 X_3 Y_4 \\ - Z_1 I_2 Y_3 X_4 + I_1 Z_2 Y_3 X_4),
\end{multline}
does. This may be seen by performing a reverse Jordan-Wigner transformation on \Qtiled, which yields
\begin{multline}\label{rev-JW-ctiled}
\Qtiledmm \xleftarrow{JW} \frac{1}{2} ( -a_3^\dagger a_1^\dagger a_4 a_1 + a_3^\dagger a_2^\dagger a_4 a_2 \\ + a_4^\dagger a_1^\dagger a_3 a_1 - a_4^\dagger a_2^\dagger a_3 a_2).
\end{multline}
Each second-quantized term in ~\eqref{rev-JW-ctiled} contains an equal number of creation and annihilation operators, showing that it conserves charge. Equivalently, each operator in \Qtiled commutes with the number operator $\sum_i a_i^\dagger a_i$, showing that the former is charge-conserving. 

It should be pointed out that constructing charge-conserving operators from a particular set of Paulis is not unique. For example, consider the same Paulis as before, $Z_1 I_2 X_3 Y_4, \: I_1 Z_2 X_3 Y_4, \: Z_1 I_2 Y_3 X_4,\: I_1 Z_2 Y_3 X_4$. While one implementation of a charge-conserving operator is shown in Eq.~\ref{charge-conserving-1}, the operators 
\begin{align*}
    \Qtiledmm_1 &= \frac{1}{2} (Z_1 I_2 X_3 Y_4 - Z_1 I_2 Y_3 X_4), \\ \Qtiledmm_2 &= \frac{1}{2}(I_1 Z_2 X_3 Y_4 - I_1 Z_2 Y_3 X_4),
\end{align*}
also conserve charge. Performing reverse Jordan-Wigner transformations on these gives
\begin{align*}\label{rev-JW-ctiled2}
\Qtiledmm_1 &\xleftarrow{JW} (a_4^\dagger a_3-a_3^\dagger a_4) - 2(a_3^\dagger a_1^\dagger a_4 a_1 + a_4^\dagger a_1^\dagger a_3 a_1), \\
\Qtiledmm_2 &\xleftarrow{JW} (a_4^\dagger a_3 - a_3^\dagger a_4) - 2(a_3^\dagger a_2^\dagger a_4 a_2 + a_4^\dagger a_2^\dagger a_3 a_2),
\end{align*}
showing that the charge-conserving constructions are not unique. Thus, there are various choices of operators that could be used. Here, we consider constructions of the form Eq.~\ref{charge-conserving-1}. The full set of operators in the \Qtiled pool is available in the \href{https://github.com/KarunyaShirali/BreakingSymmetries}{GitHub repository} containing the data that support the findings of this study. 

Finally, an alternative method to perform operator selection for operator pool tiling is to use the qubit-excitation-based~\cite{Yordanov2021} operator pool when running \Adapt on the small problem instance. This procedure is not entirely problem-agnostic, since the candidate operators in the QEB pool are all charge-conserving. In addition, tiling the chosen operators here simply generates a subset of the full QEB pool, already considered here as the \xQx \: pool (when ``double'' qubit-excitations are not included in the QEB pool). Therefore, we focus on using the Pauli pool in our simulations.

\subsubsection{The translation-invariant Pauli Pool}
\label{sec:TIPP}

Individual Pauli operators break translation invariance. In order to investigate the effect of translation invariance, we also construct translation-invariant versions of each Pauli operator to form the \Atiled (read as $\Lambda$-tiled) pool. We note that we define two distinct types of translation-invariant operators here: those that connect fermionic staggered sites (`even'-index sites) with each other, and those that connect anti-fermionic sites (`odd'-index sites). Translationally-invariant sums of terms (`volume' operators) are accompanied by `surface' operators that act only on the boundaries.

For example, the operator 
\begin{eqnarray*}
   \Paulimm_i &=& [XYZZ]_i
   \ \equiv\ X_i Y_{i+1} Z_{i+2} Z_{i+3} 
\end{eqnarray*}
of tile size $2L_{\rm tile}=4$ leads to the following \Atiled operators,
\begin{eqnarray*}
    \Atiledmm_{Vp} &=& \sum_{i=0,2,\ldots}^{N} \Paulimm_{i+p}
    \qquad (p \in \{1\equiv{\rm odd}, 2\equiv{\rm even}\})
    \\
    \Atiledmm_{Sp} &=& \Paulimm_p + \Paulimm_{N-(2L_{\rm tile}-2)-p}
    \qquad (p \in \{1,2\}),
\end{eqnarray*}
where the summation in the first operator runs until the end of the lattice of size $N\equiv 2L$ is reached, in the sense that the last tile included is the last one
that still fits within the system.
We do not translate the operator over the boundaries, keeping in mind the open boundary conditions.

\subsection{Summary of Pools}
Below we list the pools used in this paper, and identify how they relate to pools in existing literature.

\begin{itemize}
\item \AQZ: translation-invariant, charge-conserving single-fermionic-excitation operators first presented in \cite{Farrell:2023fgd}.
\item \AQx: translation-invariant, charge-conserving single-qubit-excitation operators (alternatively, translational-invariant version of the qubit-excitation~\cite{Yordanov2021} pool, with ad hoc surface terms).
\item \AxZ: translation-invariant operators obtained by breaking each \AQZ operator into two non-charge-conserving terms.
\item \xQZ: single-fermionic-excitation operators.
\item \xxZ: obtained by breaking each \xQZ operator into individual Pauli terms.
\item \xQx: single-qubit-excitation~\cite{Yordanov2021} operators, without the ``double-qubit-excitation'' terms.
\item \Axx: translational-invariant version of the qubit-ADAPT~\cite{Tang2021} pool, with ad hoc surface terms.
\item \xxx: qubit pool~\cite{Tang2021}, but without the ``doubles'' terms.
\item $\Paulimm$: an operator pool constructed by embedding the chosen operators $\Paulimm_j$ local to $L_{\rm tile}$ into an identity string on the full system, yielding operators of the form $I^{\otimes i}\otimes \Paulimm_j \otimes I^{\otimes (n-i-L_{\rm tile})}$.
\item Full Pauli pool: a pool consisting of all possible Pauli strings on $n$ qubits, used within the procedure in Sec.~\ref{operator_tiling}.
\item \Qtiled: charge-conserving operators obtained by combining commuting chosen operators $\Paulimm_j$.
\item \Atiled: translation-invariant versions of the chosen operators $\Paulimm_j$.
\end{itemize}

\section{Results}
\label{sec:results}
In our simulations, we prepare the ground-state of the lattice Schwinger Hamiltonian (Eq.~\ref{eq:schwingerham}) on system sizes from $L=2$ to $10$, i.e. $4$ to $20$ staggered sites. We focus on three different points in parameter space of the model:
\begin{itemize}
  \item[A)] 
    $m_0=0.5$, $g=0.3$
  \item[B)] 
    $m_0=0.1$, $g=0.8$ 
  \item[C)] 
    $m_0=0.1$, $g=0.3$\ .
\end{itemize}

These values are chosen as in~\cite{Farrell:2023fgd} to explore a range of correlation lengths $\xi$ in the model, with \easy $<$ \medium $<$ \hard.

Henceforth, we label models by their correlation length.
In the main text, we mainly focus on the model \hard; Appendix~\ref{easy-medium} gives results for the models \easy and \medium.
The initial state is chosen to be $|\psi_{\REF}\rangle = |10\rangle^{\otimes L}$. This state preserves the translation-symmetry, is in the correct charge sector ($Q=0$), and also preserves time-reversal symmetry. 
For the tiling experiments, we use a tile of size $L_{\rm tile}=2$, i.e. 4 staggered sites.
The gradient convergence threshold for \Adapt is set to be $\varepsilon=10^{-3}$. We also make use of TETRIS-\Adapt~\cite{Anastasiou2024:TETRIS}, in which multiple disjoint operators are added in batches at each step. We use the BFGS algorithm as implemented in the \verb|Optim.jl| Julia software package \cite{julia,optim} with a parameter gradient convergence criterion of $10^{-6}$ to optimize the variational coefficients. 

\begin{figure}    
    \includegraphics[width=\columnwidth]{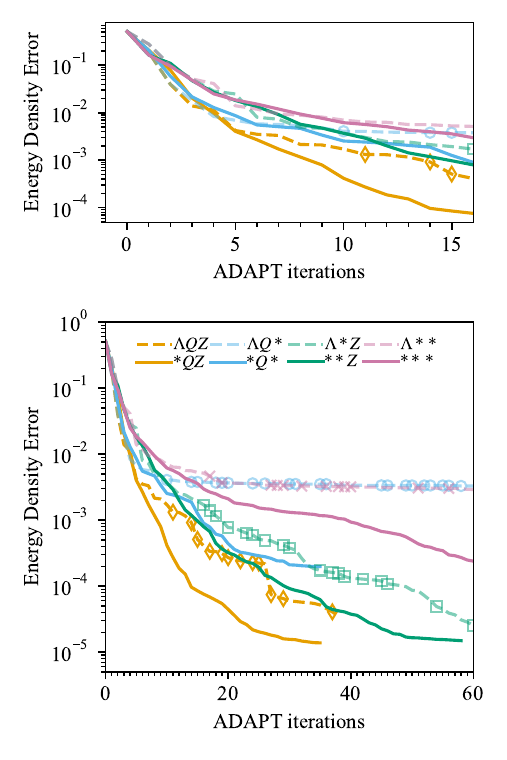}
    \caption{Evolution of the ADAPT energy density error for $L=9$ with respect to the number of ADAPT iterations for \hard. The markers indicate where the surface operators are selected for the \A-pools. The top panel is an inset of the full optimization for $L=9$ which shows the trajectories in the initial stage of the algorithm, and where the \textit{first} surface operators are selected.}
    \label{L9_energyerror_adapt}
\end{figure}

\subsection{Relaxing symmetries}\label{relaxing-symmetries}
The first measure that we use to assess the pools is the energy density error $(E_{\text{ADAPT}} - E_0)/L$, where $E_0$ is the exact ground-state energy. 
The results of relaxing the coordinate invariance \A, charge conservation \Q and fermionic 2-locality \Z in the operator pools on the energy density error are shown in Fig.~\ref{L9_energyerror_adapt} for a representative system size $L=9$ (we note that the difference in performance between the various pools becomes more visible as the correlation length increases, see Appendix~\ref{easy-medium} for the results for the systems with lower correlation lengths (\easy and \medium). The markers show where the surface operators are selected for the \A-pools.
Two features become apparent: first, the \Z-pools consistently reach the lowest energy density errors in the fewest \Adapt iterations, while the non-\Z-pools \xQx, \xxx, \AQx and \Axx pools are less accurate. Second, most operators selected for the \A-pools are surface operators. Since we would like to gauge the pools' effectiveness when scaling to larger systems (in which the surface effects become less significant), we focus on the regions \textit{before} the surface operators are picked. 

The top panel in Fig.~\ref{L9_energyerror_adapt} shows the trajectories in the initial stage of the algorithm. The markers show where the first few surface operators are selected for the \A-pools and they serve as an indicator of the accuracies the \A-pools achieve before the boundary effects come into play. We find that while all the \A-pools initially lower the energy, there is a change in their behavior between iterations ${\sim 10-15}$ where \AQx and \Axx plateau. Notably, the region where they start to flatten is also where the first surface operators are selected for \AQx and \Axx. This suggests that while the \A-pools are relevant for the bulk, the boundary effects overshadow their performance fairly soon. 

In general, we find that the pools retaining \Z lead to the lowest energy error. This is observed to be due to the fact that we are preparing mean-field solutions (which have a large overlap with the ground-state in this model) with the \Z-pools, see Appendix~\ref{mean-field-reference-states} for more details.

\begin{figure}
    \includegraphics[width=\columnwidth]{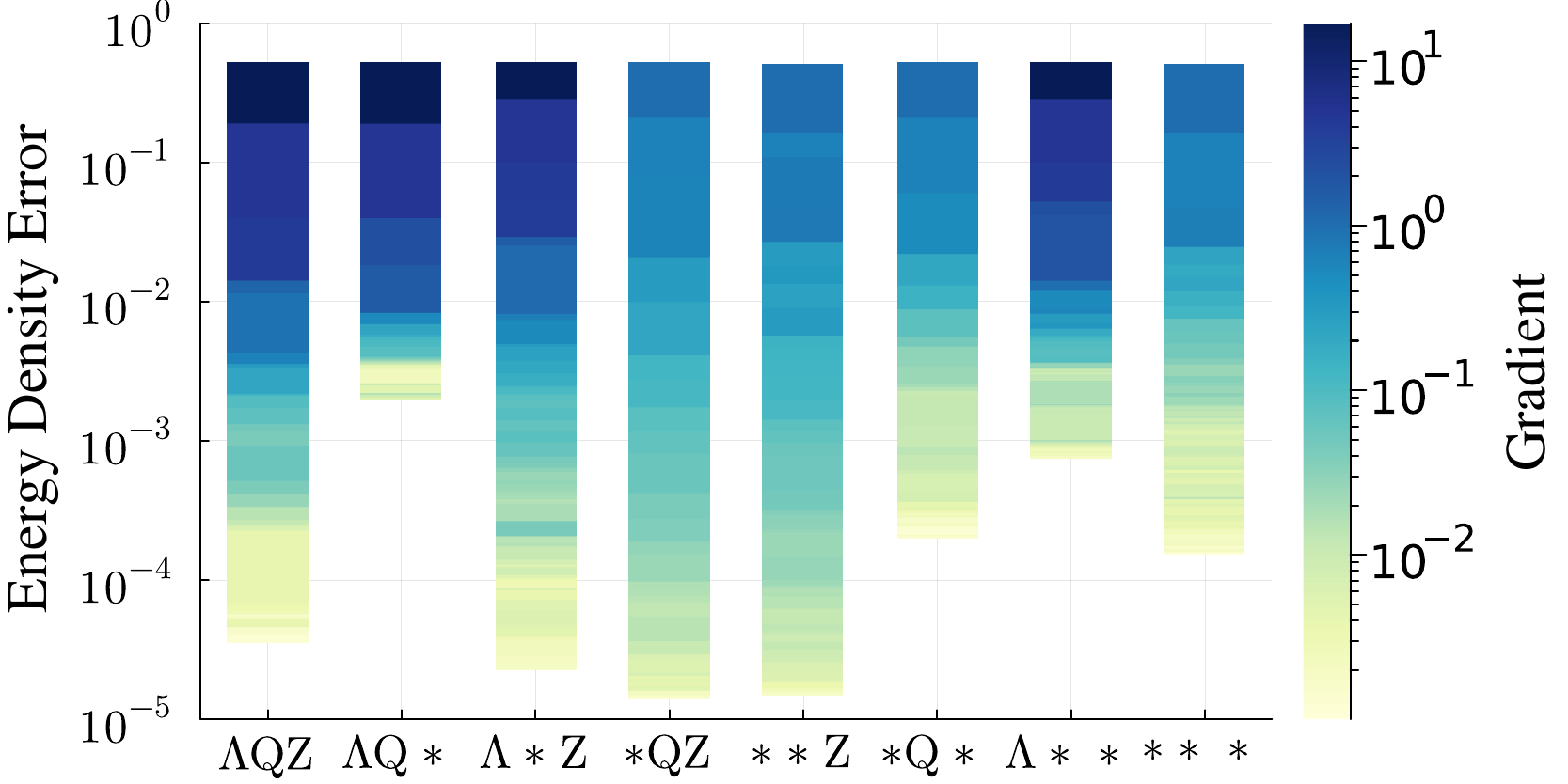}
    \caption{Energy density error evolution versus largest pool gradient during \Adapt for $L=9$, \hard. The gradient magnitudes are shown using the color scale on the right. }
    \label{fig:scores}
\end{figure}

In Fig.~\ref{fig:scores}, we plot the magnitude of the largest pool gradient (i.e., the gradient of the first of the disjoint operators added to the ansatz at that step) at each \Adapt iteration. The gradient magnitudes are represented by the color bar on the right.
A healthy convergence trajectory usually involves the energy density error decreasing in proportion with the gradients; ideally, the pool gradients should become small only near the energy minima.
We find that the change in energy density error for a given decrease in gradient magnitude is greatest for the \xQZ, \xxZ pools, followed by the \AxZ, \AQZ, \xQx and \xxx pools. The \xQx and \xxx pools do not reach the energy accuracy that \xQZ and \xxZ yield because they reach the gradient convergence criterion sooner (see Fig.~\ref{L9_energyerror_adapt}).
The gradients of \Axx and \AQx fluctuate back and forth with relatively large magnitudes (compared to the convergence threshold $10^{-3}$). These unfavorable fluctuations are likely due to finite-size effects in which the algorithm picks surface operators that correct the boundary; while having non-zero gradients, these operators are not effective at optimizing the bulk.  

\begin{figure}
    \centering
    \includegraphics[width=0.9\columnwidth]{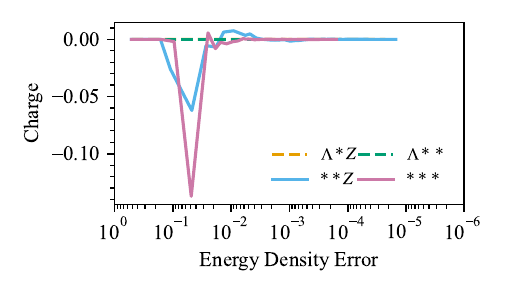}
    \caption{Charge variation in the ans\"atze as energy density error decreases during \Adapt for $L=9$. }
    \label{fig:charge-variation}
\end{figure}

In Fig.~\ref{fig:charge-variation} we plot the deviations in charge from $Q=0$ in the ansatz for the charge non-conserving pools as the \Adapt algorithm proceeds, in order to assess its impact on lowering the energy. We find that the largest charge variations happen in the early part of the algorithm, and that the charge automatically goes back to $0$ as the energy density error falls below $10^{-3}$. This reveals that it is energetically advantageous to explore different subspaces early on; for example, the charge of the \xxx pool reaches as much as $14\%$ of the charge of the first excited state. We find that the ansatz for the \xxx pool at this stage explores different charge sectors, with the largest probability outside of the charge-zero sector coming from the $Q=-2$ and $Q=2$ sectors. 
Later on, however, the algorithm chooses not to leave the $Q=0$ subspace.

\begin{figure}
    \centering
    \includegraphics[width=\linewidth]{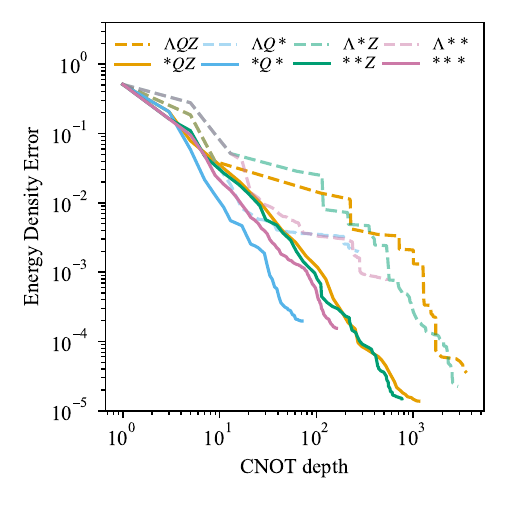}
    \caption{Energy density error variation with CNOT depths obtained by transpiling the optimized ans\"atze from each operator pool.}
    \label{fig:CNOTs}
\end{figure}

We quantify the quantum resource requirements by comparing the CNOT depths of the optimized ans\"{a}tze from each pool calculated using the qiskit~\cite{qiskit2024} transpiler in Fig.~\ref{fig:CNOTs}. The CNOT depth is a major bottleneck in calculations run on current quantum hardware, and finding compact circuits with lower CNOT resource requirements that simultaneously yield high-fidelity representations of the ground-state is desirable. The $Z$-strings in the Jordan-Wigner transformation typically increase the CNOT overhead significantly. We find that the the \xQZ and \xxZ pools reach the lowest energy density error, with similar CNOT depths. The \AQZ and \AxZ pools also converge to energies close to those of \xQZ and \xxZ, but with considerably greater CNOT depths. The more hardware-efficient \xQx and \xxx pools have the lowest CNOT resource requirements, but converge to less accurate energies. Finally, the \{\AQx, \ \Axx\} pools naturally have a lower CNOT overhead than \{ \AQZ,\ \AxZ \} because of the absence of \Z, but terminate earlier.

\begin{figure}[!th]
    \centering
    \includegraphics[width=\linewidth]{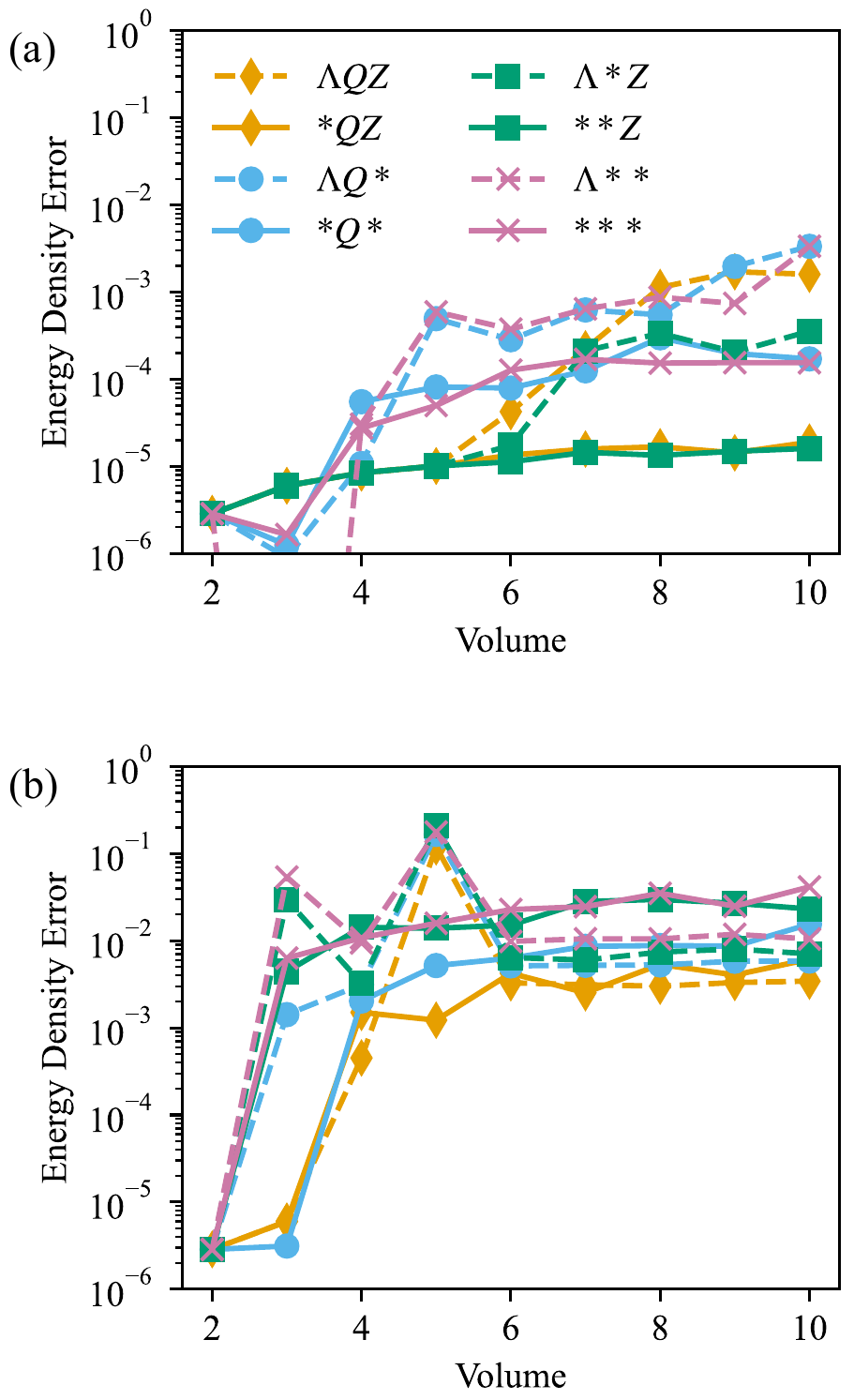}
    \caption{ADAPT energy density errors achieved for $L=2-10$ obtained by imposing budgets on quantum resources. The results shown are for
        (a) when a budget of $1000$ CNOT gates is used, and
        (b) when a budget of $100$ function evaluations is used.
    }
    \label{fig:budget}
\end{figure}

In Fig.~\ref{fig:budget} we show the results of assuming finite budgets for the number of CNOT gates and function evaluations on the energy density error, for the different system volumes considered. We focus especially on the system volumes $L=5-10$, since they display trends that are expected to be relevant for larger system volumes. We find that the \A-pools significantly increase the CNOT gate overhead, and achieve less accurate energies compared to the best-performing \{ \xQZ,\xxZ \} pools, for a CNOT cutoff of $1000$. On the other hand, when considering a budget of $100$ function evaluations, the \A-pools perform better than the non-\A pools (barring \xQZ). This suggests that retaining translation invariance is beneficial for the optimization sub-routine.

\subsection{Operator tiling}
\begin{figure}
    \includegraphics[width=\linewidth]{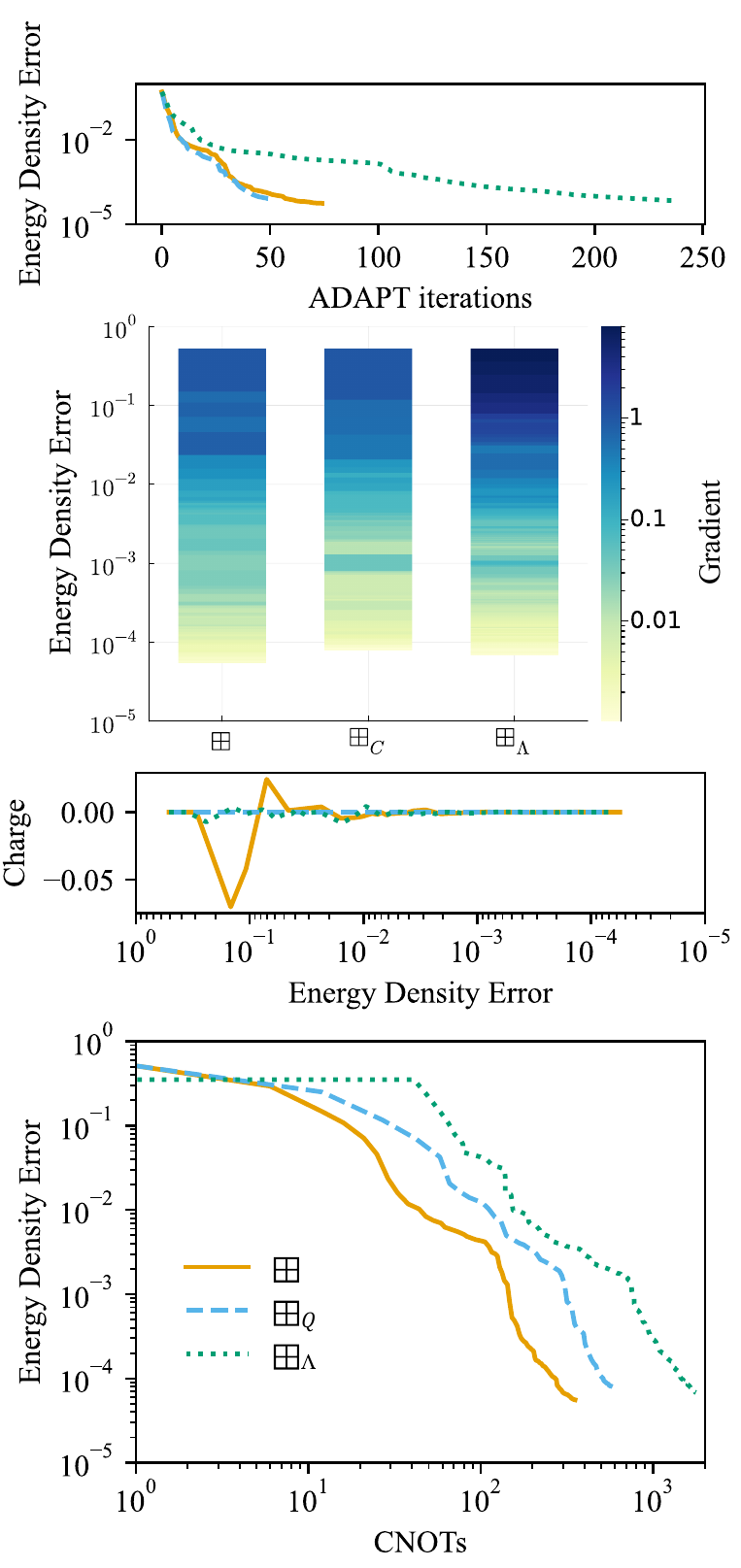}
    \caption{Results of running \Adapt using the tiled operator pools for $L=9$.
        (a) Evolution of the \Adapt energy density error with respect to the number of ADAPT iterations.
        (b) Change in energy density error with largest pool gradient.
        (c) Change in the charge $Q$ as the algorithm progresses.
        (d) Energy density error variation with CNOT depths obtained by transpiling the optimized ans\"atze from each operator pool.
    }
    \label{fig:tiled}
\end{figure}

Next, we present the results of using the tiled Pauli pool ($\Paulimm$), the \Qtiled pool, and the \Atiled pool. Upon performing the operator selection step ~\ref{degenerateADAPT} in Section~\ref{operator_tiling}, we find that the operators chosen by \Adapt preserve the fermion/anti-fermion number parity, as well as time-reversal symmetry. We then construct the \Pauli, \Qtiled and \Atiled pools: Fig.~\ref{fig:tiled} shows how the three compare. We find that the \Qtiled pool generally converges in fewer \Adapt iterations than the other two, and all three achieve similar final energies. In terms of the gradients, we observe that the energy density errors decrease at a similar rate with respect to the largest pool gradient for all three pools, as plotted in Fig.~\ref{fig:tiled}(b). The charge evolution during the \Adapt algorithm is shown in Fig.~\ref{fig:tiled}(c); we find that the algorithm finds it energetically favorable to leave the $Q=0$ subspace in the initial steps when using the \Pauli pool, but returns to the subspace as it proceeds, similar to the top-down pools. Finally, upon transpiling to one- and two-qubit gates, we observe that the tiled Pauli pool \Pauli achieves better energy accuracy with the lowest CNOT depth, as shown in Fig.~\ref{fig:tiled}(d).

Our results from the tiled pools indicate that, while conserving charge \Q leads to shorter \Adapt trajectories (and hence, optimization complexity), the circuit CNOT depths do not differ significantly from the other two tiled pools. We also note that the construction of the \Qtiled pool is not unique; it might be interesting to compare different types of charge-conserving pools in the future.

\subsection{Time-reversal symmetry breaking}\label{TRS-breaking}
In this section, we discuss the results of experiments in which time-reversal symmetry is broken in the reference state by introducing complex amplitudes. In our experiments, we `contaminate' the reference state by introducing into it complex amplitudes. 
For $L=4$, for example, we define the \T-breaking reference state as:
\begin{equation}
    |\psi_1\rangle = \frac{1}{\sqrt{2}} | 10101010\rangle -\iu |10110010\rangle,
\end{equation}
as well as the \T-preserving reference state as:
\begin{equation}
    |\psi_2\rangle = \frac{1}{\sqrt{2}}( | 10101010\rangle - |10110010\rangle).
\end{equation}
Note that the states $|\psi_1\rangle$ and $|\psi_2\rangle$ have charge $Q=0$. For the \T-relaxing operator pool, we use a charge (\Q) and \Z-preserving operator pool consisting of two kinds of operators: those that are of the form in Eq.~\ref{eq:C_op} along with those of the form in Eq.~\ref{eq:C_op:T}. For the \T-preserving experiments, the operator pool only contains operators of the form in Eq.~\ref{eq:C_op}.

We specify a parameter to quantify the \T-breaking, defined as
\begin{equation}
    \Delta_{\rm T}(\psi) = |\text{Im}(\psi)|/|\text{Re}(\psi)|,
\end{equation}
where the global phase of $\psi$ is selected to minimize $|\text{Im}(\psi)|$ before calculating $\Delta_{\rm T}$.

\begin{figure}
    \centering
    \includegraphics[width=0.9\linewidth]{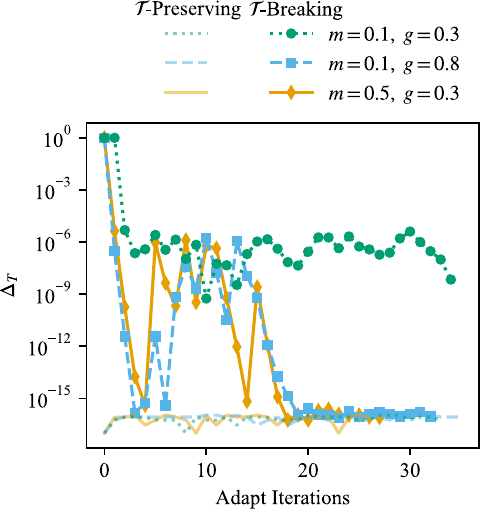}
    \caption{Time-reversal breaking parameter as a function of \Adapt iterations for $L=9$. }
    \label{fig:TRS-breaking}
\end{figure}

In Fig.~\ref{fig:TRS-breaking}, we plot the extent to which \T is broken as \Adapt progresses. The lighter-shade data shows the results of the \T-preserving experiment, which serves as a control, and the stronger-shade data shows the results of \T-breaking. We find that the only \T-breaking operator ever chosen is at the first (\easy, \medium) or second (\hard) ADAPT step; this means that \Adapt chooses to restore \T very quickly in all cases. Recalling that multiple operators are added at each step~\cite{Anastasiou2024:TETRIS}, we observe that for \easy, the \T-breaking operator has the largest gradient. On the other hand, for \medium, the \T-breaking operator chosen has the smallest gradient among the operators chosen, with the rest of the operators all having the same (degenerate) gradient. For \hard, the \T-breaking operator has a gradient that lies between the minimum and maximum of the pool gradients. This indicates that there are different energy scales operating in the problem. For \medium and \hard, there is a larger energy scale that wants to `fix' the other parts of the lattice rather than immediately restore \T, whereas for \easy, restoring \T gives rise to the greatest energy decrease initially. We also note that when gradients are degenerate, \Adapt selects the operator to add arbitrarily. This suggests that, especially for \medium and \hard, where \T-breaking does not have the largest energy scale initially, a different choice of operator selection resulting from the degenerate gradients could have resulted in a slightly different trajectory. The qubit support of the (arbitrarily) chosen operator affects the choice of operators to add next - recall that the algorithm selects \emph{disjoint} operators in decreasing order of gradient magnitude - which would determine whether the \T-restoring operator is selected in the first iteration, or later when the energy scales are closer. However, we expect \T to be restored within the first few iterations in most cases.   

\section{Discussion}
\label{sec:discussion}
Keeping various factors in mind, including the number of steps to convergence, and CNOT depths, we find, overall, that breaking the translation-symmetry \A, but preserving charge conservation \Q are beneficial for the lattice Schwinger model at the system volumes we study. This is because most of the \A-pool calculations are found to be focused on correcting for the boundary effects rather than optimizing the bulk. Consequently, we predict the \A-pools to become more useful for larger systems where the boundary effects are less significant. 

Next, we find that there is no noticeable benefit to relaxing charge conservation \Q in the pools. Keeping the limitations of current quantum hardware in mind, we note that the \Q-preserving pools require fewer \Adapt iterations than their \Q-breaking partners to converge, which lead to fewer function evaluations overall. The circuit depths of the resultant ans\"atze from each pair of \Q-preserving and \Q-breaking pools are similar. This leads us to conclude that preserving \Q is favorable. 

Retaining the Jordan-Wigner $Z$-strings \Z is also found to be beneficial. It should be noted that this finding is influenced by the fact that the ground-state of the lattice Schwinger model has a large overlap with the mean-field solution.

Finally, we find that the algorithm is strongly sensitive and averse to breaking time-reversal symmetry: the algorithm finds that it is energetically favorable to quickly restore \T. The differences in the gradients of the \T-breaking operators, however, between \easy, \medium and \hard suggest that there might be Hamiltonians in which other energy scales become more relevant and the restoring of \T happens more gradually.

Assessing the performance of an operator pool requires taking into consideration the finite coherence times of qubits, the errors that limit quantum gate fidelities, as well as the measurement errors that arise from the limited number of shots that can be performed on current hardware. In general, shallow-depth circuits that converge in fewer iterations are preferable. Shallow circuits are beneficial because the time to perform them is short and there are likely to be fewer gate errors; convergence in fewer iterations is preferable because it leads to lower optimization complexity and hence, fewer evaluations of the objective function. 
Here, we find that breaking translation invariance is preferred when considering the CNOT depths of the resultant circuits. For current noisy quantum devices, we expect that the circuit depths will be the limiting factor in simulations of lattice models, thus supporting breaking translation invariance. Now, an argument against the use of translation-invariant operators for this model is that the open boundary conditions of the Schwinger lattice, which are necessary in order to integrate out the gauge field, could favor relaxing translation invariance. Indeed, we observe that the \A-pools are effective at decreasing the energy before the boundary effects come into play. However, we expect the boundaries to become less important as $L\rightarrow\infty$. 
It is thus possible, for example, that the \A-pools, especially the relatively more hardware-efficient \AQx and \Axx pools, can reach accurate energies, without resulting in prohibitively deep circuits, for large system volumes that are beyond the capacity of our computational resources. Strategies such as including \Z-preserving operators only at the surfaces (discussed in Appendix~\ref{sec:modified_surface}) to account for finite sizes might also be more tractable in terms of quantum resources, since the increase in CNOT overhead from such operators localized near the boundaries will likely not be significant. 
We also find that assuming a limited budget for the number of objective function evaluations is found to favor \emph{preserving} translation invariance. This suggests the use of translation-invariant pools for lattice model simulations on future, error-corrected quantum devices, especially so for system volumes in which the finite-size effects are negligible.

Additional studies on truly periodic lattices could be used to further probe the importance of translation-symmetry. Retaining translation-symmetry in the ansatz has various benefits: for instance, measurements on a translation-invariant wavefunction can be performed highly efficiently, and such ans\"{a}tze are amenable to extrapolation to large volumes~\cite{Farrell:2023fgd,Gustafson2024:sc2}. It might also be interesting to take the union of all pools defined here to create a pool that would have greater flexibility without increasing the measurement costs. For example, measuring the gradients of the \AQZ pool gives access to the gradients of all the \Z-pools, since they are implemented using the same Pauli strings. Similarly, the \Qtiled and \Atiled pools are constructed using the tiled \Pauli operators, meaning that the former two pools' gradients can be obtained based on the gradients of the \Pauli pool, without performing additional measurements.

\section*{Data Availability}
The code and data that support the findings of this study are openly available at \url{https://github.com/KarunyaShirali/BreakingSymmetries}.

\acknowledgments
The authors thank Nicholas Mayhall, Bharath Sambasivam and David Frenklakh for helpful comments. K. Shirali, K. Sherbert, A.F., Y.C., A. W., S.E.E., R.D.P., 
are supported by the U.S. Department of Energy, Office of Science, National Quantum Information Science Research Centers, Co-design Center for Quantum Advantage (C$^2$QA) under Contract No. DE-SC0012704. 
A.W. was supported by the U.S. Department of Energy, Office of Science, Basic Energy Sciences, Materials Sciences and Engineering Division. 
A.F. is supported by the DFG through the Emmy Noether Programme (project number 545261797).

The authors acknowledge Advanced Research Computing at Virginia Tech for providing computational resources and technical support that have contributed to the results reported within this paper. URL: \url{https://arc.vt.edu/}.

\appendix
\section{Results for other parameter points}\label{easy-medium}

\begin{figure}
    \includegraphics[width=\linewidth]{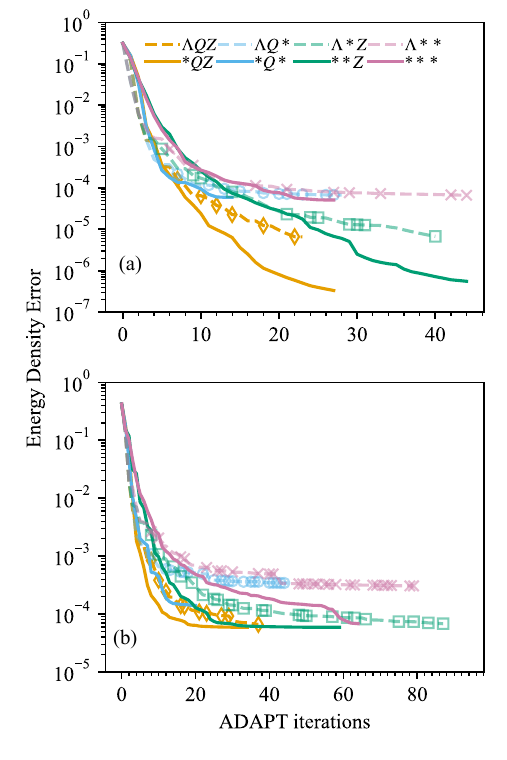}
    \caption{Evolution of the \Adapt energy density error for $L=9$ 
     with respect to the number of \Adapt iterations for
        (a) \easy, and
        (b) \medium.
    }
    \label{fig:appendix_L9_energyerror_adapt}
\end{figure}

The results for running \Adapt for systems \easy and \medium, $L=9$, are shown in Fig.~\ref{fig:appendix_L9_energyerror_adapt}. In Fig.~\ref{fig:appendix_L9_energyerror_adapt}(a) (\easy) we find that the \xQZ and \xxZ pools reach the lowest energy error, followed by \AQZ and \AxZ; the remaining four pools all reach similar final energies with errors on the order $10^{-4}$. In Fig.~\ref{fig:appendix_L9_energyerror_adapt}(b) (\medium), the differences between the pools are \emph{less} pronounced than for \easy. This is due to the fact that the mean-field solution for \medium has a lower overlap with the exact ground-state than \easy (see Fig.~\ref{fig:meanFieldSoln}). Thus, the \Z-pools, which can at best prepare the mean-field solution (given that they are derived from fermionic single-excitations), achieve less accurate final energies, and their distinction from the non-\Z pools is suppressed.

\section{Modifying the surface operators}\label{sec:modified_surface}

\begin{figure}
    \subfloat[\label{surface_noZ}]{
        \includegraphics[width=0.8\linewidth]{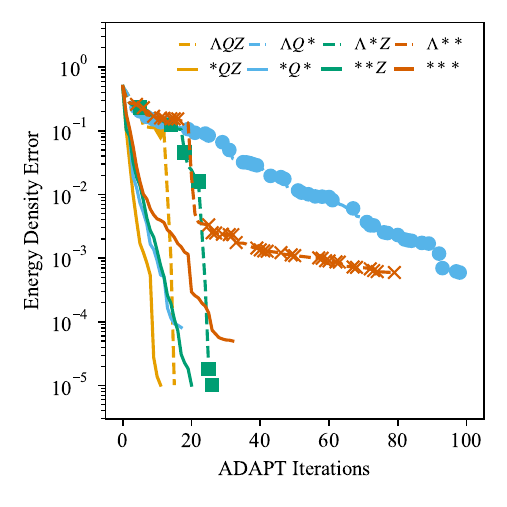}
    }
    \hfill
    \subfloat[\label{surface_Z}]{
        \includegraphics[width=0.8\linewidth]{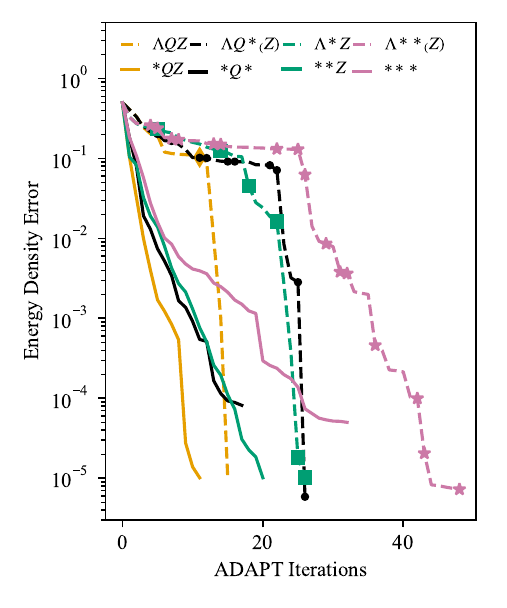}
    }
    \caption{Evolution of the ADAPT energy density error for $L=5$ with respect to the number of ADAPT iterations for \hard. The markers indicate where the surface operators are selected for the \A-pools.
        (a) Using the pools as defined, and
        (b) when \Z surface operators replace the non-\Z surface operators in the \AQx and \Axx pools.
    }
    \label{fig:surfaceZ}
\end{figure}

In Sec.~\ref{relaxing-symmetries}, we found that the \AQx and \Axx pools' trajectories plateau after the surface operators start being chosen, which could indicate that the surface operators are not sufficiently expressive to `fix' the surface. On the other hand, \AQZ and \AxZ do not experience the unfavorable plateauing effect when their surface operators are chosen, suggesting that the $Z$-strings are especially helpful near the boundaries. In order to investigate this further, we run simulations in which we replace the surface operators in \AQx and \Axx  with surface operators that include \Z. The results are shown in Fig.~\ref{fig:surfaceZ} for a representative system $L=5$, where Fig.~\ref{fig:surfaceZ}(a) uses the pools as originally defined and Fig.~\ref{fig:surfaceZ}(b) modifies the surface operators in \AQx and \Axx. We find, as shown in Fig.~\ref{fig:surfaceZ}(b), that the modified \AQx and \Axx pools now achieve similar accuracies to the best-performing \xQZ and \xxZ pools, but the \A-pools still have longer trajectories than non-\A. 

\begin{figure}
    \subfloat[\label{fig:error_L_noZ}]{
        \includegraphics[width=0.8\linewidth]{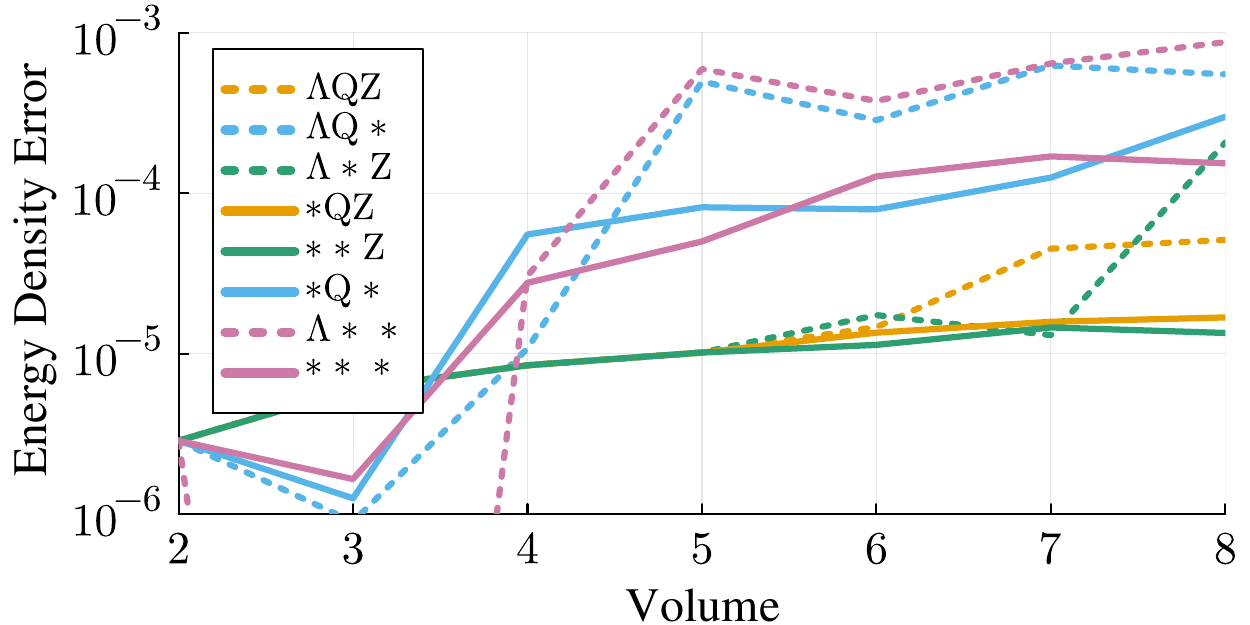}
    }
    \hfill
    \subfloat[\label{fig:error_L_withZ}]{
        \includegraphics[width=0.8\linewidth]{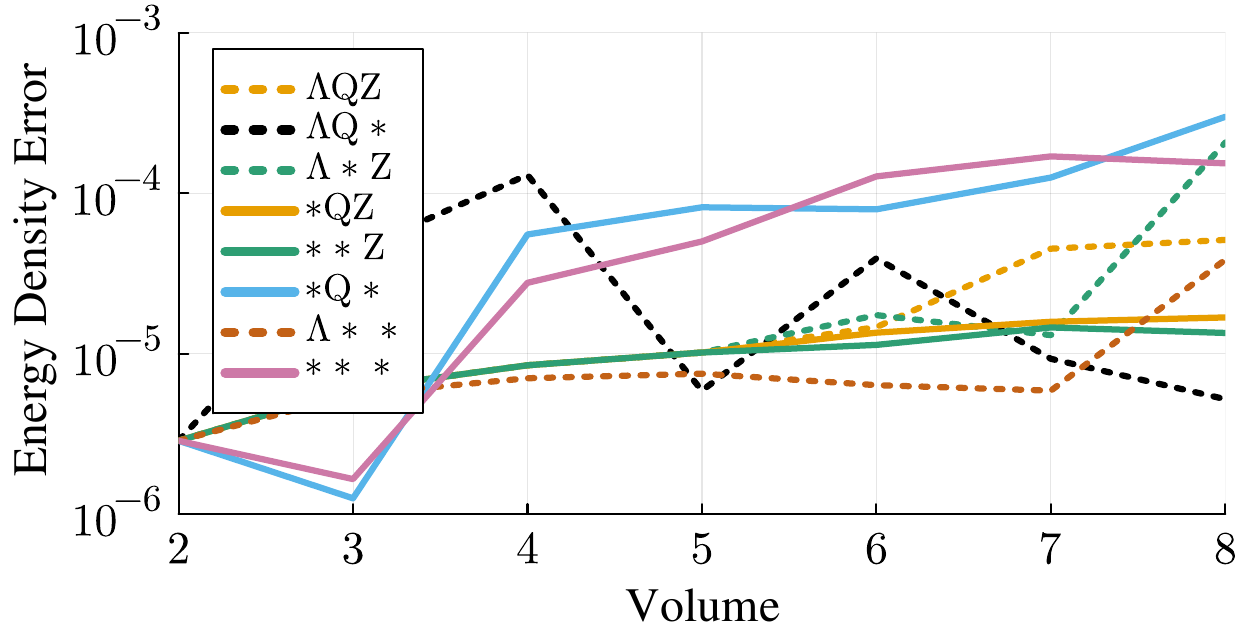}
    }
    \caption{Evolution of the ADAPT energy density error for system volumes $L=2-8$, for a CNOT depth cutoff of $1500$. 
        (a) Using the pools as defined, and
        (b) when \Z surface operators replace the non-\Z surface operators in the \AQx and \Axx pools. 
    The operators in the other pools remain unchanged.
    }
    \label{fig:error_L_surfaceZ}
\end{figure}

In Fig.~\ref{fig:error_L_surfaceZ} we show the results of cutting off the solutions at a CNOT depth of $1500$ for system volumes $L=2-8$: Fig.~\ref{fig:error_L_surfaceZ}(a) uses the original pool definitions and Fig.~\ref{fig:error_L_surfaceZ}(b) swaps the surface operators in \AQx and \Axx with surface operators retaining \Z. We find that modifying the surface operators in this manner improves the energy accuracy, without significantly increasing the CNOT overhead. This suggests that it might be a tractable strategy for simulations of larger volumes on devices in which the circuit depths are limiting factors.

\section{Mean-field solutions as reference states}\label{mean-field-reference-states}
As discussed in the manuscript, the ``parent pool'' \AQZ is built from fermionic single-excitation operators. The derivative pools \xQZ, \AxZ, and \xxZ (in which \Z is retained) thus also correspond strictly to one-body excitations or linear combinations of them. Dropping the Jordan-Wigner $Z$-strings, however, results in higher-body excitations being included in the operators. Bearing this in mind, we note that the \AQZ, \xQZ, \AxZ, and \xxZ pools can, at best, prepare the mean-field (in other words, Hartree-Fock) solutions of the model. The mean-field solution is one in which the wave function is optimized under one-body excitations to achieve the lowest energy \cite{Thouless_1960_NP,Google_2020_Science}.
Conversely, the \AQx, \xQx, \Axx and \xxx pools are not strictly one-body operators, and do not have such a connection to the mean-field solution. 

\begin{figure}
    \centering
    \includegraphics[width=\linewidth]{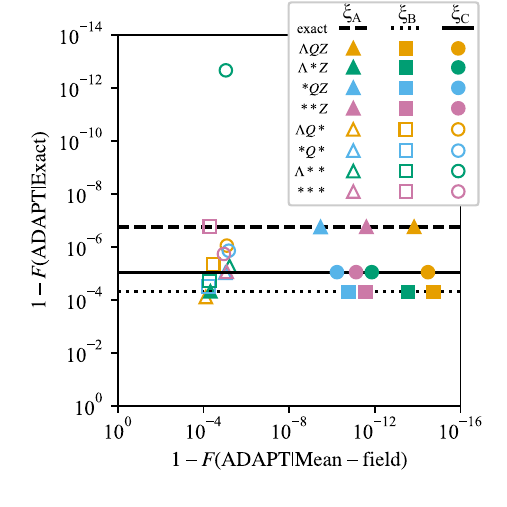}
    \caption{Mean-field solutions for \easy, \medium and \hard, for $L=3$.}
    \label{fig:meanFieldSoln}
\end{figure}

In Fig~\ref{fig:meanFieldSoln}, we plot the distances of the \Adapt wave functions to the exact solutions against those of the mean-field solutions. The \Z-pools are represented with filled markers, whereas the non-\Z pools are represented with unfilled markers. 
The three horizontal lines serve as a reference to show the infidelities between the exact ground-state and the mean-field solution for each parameter point. 
We notice immediately from the horizontal reference lines that the exact solutions for each parameter point have a large overlap (infidelity $< 10^{-4}$) with the corresponding mean-field solutions. Next, we observe that the optimized ans\"atze for the \Z-pools (filled markers) are found to have high fidelities with the mean-field solution, whereas the non-\Z pools do not, reinforcing the fact that the \Z-pools prepare mean-field solutions. This leads to the question: 
can the non-\Z pools achieve better representations of the ground-state than the \Z-pools, in particular, if they are started from a state that has a significant overlap with the target state (i.e., the mean-field solution)?

\begin{figure}
    \centering
    \subfloat[\label{fig:infid_meanFieldStart}]{
        \includegraphics[width=0.9\linewidth]{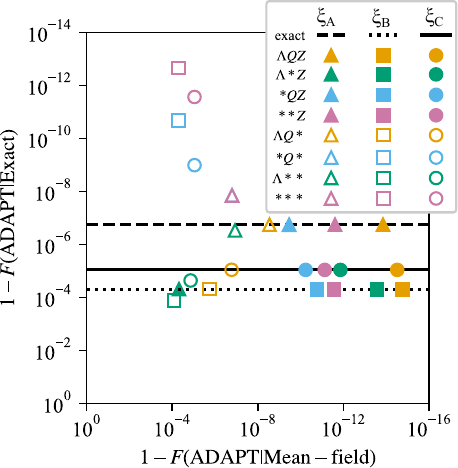}
    }
    \hfill
    \subfloat[\label{fig:energyerror_meanFieldStart}]{
        \includegraphics[width=0.9\columnwidth]{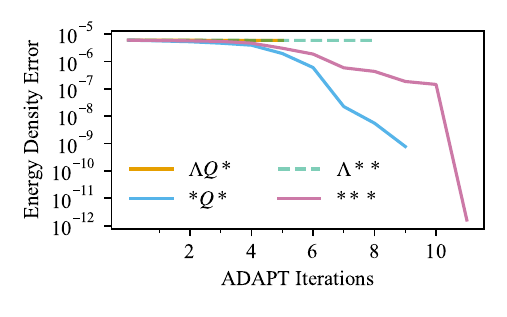}
    }
    \caption{Results obtained when \Adapt is started using the mean-field solution as the reference state for $L=3$.
        (a) infidelity between the \Adapt ans\"atze and the exact ($y$-axis), as well as mean-field solution ($x$-axis), and
        (b) energy density error. The data for \AQx is directly under that for \Axx.
    }
    \label{L3_meanFieldStart}
\end{figure}

We perform an experiment to investigate this in which we calculate the mean-field solutions for system sizes $L \in \{2,3,\cdots7\}$ and use them as the reference states $|\psi_\REF\rangle$ for the different pools. We find that simulations using the \Z pools (\AQZ, \xQZ, \AxZ, and \xxZ) do not even start; this is because the operator gradients are all identically zero, since the mean-field solution optimizes the state under all one-body excitations, and the system is already at the minimum energy possible under one-body operations. 
On the other hand, the non-\Z pools start, provided we reduce the gradient tolerance threshold below the value of $10^{-3}$ used for data presented in the main text. With a tolerance of $10^{-5}$ the \xQx and \xxx pools are able to achieve highly accurate representations of the ground-state, see Fig.~\ref{fig:infid_meanFieldStart}.
Additionally, we observe that starting from the mean-field solution is detrimental to the \Axx and \AQx pools: the algorithm takes steps forward by growing the ansatz, but does not decrease the energy (see Fig.~\ref{fig:energyerror_meanFieldStart}). This could be due to the fact that the mean-field solution is notably \textit{not} translation-invariant by default, which means that the \A-pools, which perform translation-invariant operations, are less likely to be relevant.  

In the main text, we find that retaining \Z in the pools is beneficial; this is precisely because they take the system towards the mean-field solution $|\psi_{{\rm MF}}\rangle$, which has a large overlap with the target ground-state for the Schwinger model. However, retaining \Z has the unsatisfactory effect of greatly increasing the CNOT depths of the resultant ans\"atze. The trade-off between the high accuracy and longer depths of the \Z-pools can be used to determine whether to include \Z for pools including double-excitations. We also note that Refs.~\cite{Tang2021, Yordanov2021, ramoa2024reducingresourcesrequiredadaptvqe} have proposed highly effective and efficient pools that do not preserve \Z.

\bibliography{ref}

\end{document}